\documentstyle[12pt,epsfig]{article}
\begin{document}
\def\bt{\begin{tabular}}
\def\et{\end{tabular}}
\def\bfr{\begin{flushright}}
\def\mm{\mbox{\boldmath $ }}
\def\efr{\end{flushright}}
\def\bfl{\begin{flushleft}}
\def\efl{\end{flushleft}}
\def\vs{\vspace}
\def\hs{\hspace}
\def\sta{\stackrel}
\def\pb{\parbox}
\def\bc{\begin{center}}
\def\ec{\end{center}}
\def\sp{\setlength{\parindent}{2\ccwd}}
\def\bp{\begin{picture}}
\def\ep{\end{picture}}
\def\uni{\unitlength=1mm}
\def\REF#1{\par\hangindent\parindent\indent\llap{#1\enspace}\ignorespaces}

\noindent \bc {\Large\bf Two Kinds of Iterative Solutions\\
\vspace{.1cm} for Generalized Sombrero-shaped Potential\\
 \vspace{.3cm} in $N$-dimensional Space}

\vs*{1cm} {\large  Zhao Wei-Qin$^{1,~2}$}

\vs*{1cm}

{\small \it 1. China Center of Advanced Science and Technology
(CCAST)}

{\small \it (World Lab.), P.O. Box 8730, Beijing 100080, China}

{\small \it 2. Institute of High Energy Physics, Chinese Academy of
Sciences,

P. O. Box 918(4-1), Beijing 100039, China}

\ec
\vspace{2cm}
\begin{abstract}

Based on two different iteration procedures the groundstate wave
functions and energies for N-dimensional generalized Sombrero-shaped
potentials are solved. Two kinds of trial functions for the
iteration procedure are defined. The iterative solutions are
convergent nicely to consistent results for different choices of
iteration procedures and trial functions.

\end{abstract}
\vspace{.5cm}

PACS{:~~11.10.Ef,~~03.65.Ge}\\


\newpage

{\large

\section*{\bf 1. Introduction}
\setcounter{section}{1} \setcounter{equation}{0}

Recently the generalized radially symmetric Sombrero-shaped
potential in $N$-dimensional space is proposed by R. Jackiw[1]:
$$
V(r) = \frac{1}{2} g^2 (r^2-r_0^2)^2(r^2+Ar_0^2),\eqno(1)
$$
where $r_0^4=(2+N)/3$, $g^2$ and $A$ are arbitrary constants. He
also challenged to apply the iterative method developed by R
Friedberg, T. D. Lee and W. Q. Zhao[2] to solving this problem. His
question is properly answered by the three authors in Ref.[3] for
one-dimensional case. The same problem will be solved for the
$N$-dimensional generalized Sombrero-shaped potential in this paper.
The corresponding Schroedinger equation for the groundstate radial
wave function is
$$
(-\frac{1}{2r^{2k}}\frac{d}{dr}r^{2k}\frac{d}{dr}+V(r))\psi(r)=E\psi(r)\eqno(2)
$$
with $ k=(N-1)/2$. The boundary conditions are
$$
\psi(\infty)=0~~~~~\psi'(0)=0.\eqno(3)
$$
When $g=1$ and $A=2$ the solution of the groundstate has an
analytical form as $\psi(r)=e^{-r^4/4}$ with the eigenvalue
$E_0=r_0^6$. However, for arbitrary $g$ and $A$ the groundstate wave
function has no analytical form. In the following two iterative
solutions for the groundstate of (2) are presented. To apply the
iterative methods we introduce the trial function $\phi(r)$
satisfying another Schroedinger equation
$$
(-\frac{1}{2r^{2k}}\frac{d}{dr}r^{2k}\frac{d}{dr}+V(r)-h(r))\phi(r)
=gE_0\phi(r)=(E-\Delta)\phi(r),\eqno(4)
$$
where $h(r)$ and $\Delta$ are the corrections of the potential and
the groundstate energy. Starting from this trial function
$\phi(r)$ we perform two iterative procedures. Define the exact
wave function as
$$
\psi(r)=f(r)\phi(r)=e^{-\tau(r)}\phi(r).\eqno(5)
$$
The iteration performed for $f(r)$ and $\Delta$ is named as
$f$-iteration[2] and the one performed for $\tau(r)$ and $\Delta$ is
named as $\tau$-iteration[4] in this paper. For the $f$-iteration
two iterative series of $\{f_n(r)\}$ and $\{\Delta_n^f\}$, $n=0,~1,
~\cdots$ are introduced with $f_0(r)=1$ and $\Delta_0^f=0$. For the
$\tau$-iteration two iterative series of $\{\tau_n(r)\}$ and
$\{\Delta_n^\tau\}$, $n=0,~1,~\cdots$ are defined with $\tau_0(r)=0$
and $\Delta_0^\tau=0$. The iterations for these two sets can be
performed according to the following equations. For $f$-iteration we
have[2]
$$
\Delta_n^f=\frac{\int\limits_0^{\infty}r^{2k}\phi^2(r)
h(r)f_{n-1}(r)dr}{\int\limits_0^{\infty}
r^{2k}\phi^2(r)f_{n-1}(r)dr},\eqno(6a)
$$
$$
f_n(r)=f_n(r_c)-2\int\limits_{r_c}^{r} \frac{dy}{y^{2k}\phi^2(y)}
\int\limits_{r_C}^{y}x^{2k}\phi^2(x)
(\Delta_n^f-h(x))f_{n-1}(x)dx,\eqno(6b)
$$
where $r_C$ could be chosen as $r_C=0$ or $r_C=\infty$ and the
normalization is chosen as $f_n(r_C)=1$. As for $\tau$-iteration we
have[4]
$$
\Delta_n^\tau = \frac{\int\limits_0^{\infty}r^{2k}\phi^2(r)~
(h(r)-\frac{1}{2}~(\tau'_{n-1}(r))^2) dr} {\int\limits_0^{\infty}
r^{2k}\phi^2(r)~ dr},\eqno(6c)
$$
$$
\tau'_n(r) = 2 r^{-2k}\phi^{-2}(r)\int\limits_0^{r}y^{2k}\phi^2(y)
          [(\Delta_n^\tau-h(y))+\frac{1}{2}(\tau'_{n-1}(y))^2]~dy\eqno(6d)
$$
where $\tau'(r)=\frac{d\tau}{dr}$. The detailed derivation is
summarized in Appendix. To ensure the convergency of the iterative
methods it is necessary to construct the trial function in such way
that the perturbed potential $h(r)$ is always positive (or negative)
and finite everywhere. Specially, $h(r)\rightarrow 0$ when
$r\rightarrow \infty$. In the following we construct two different
trial functions for the iteration procedures.

\section*{\bf 2. Trial Functions}
\setcounter{section}{2} \setcounter{equation}{0}

{\large \bf Trial Function I}

Introduce
$$
\phi(r)=e^{-S_0(r)}.\eqno(7)
$$
Now substituting (7) into (4) we obtain the equation for $S_0(r)$:
$$
S_0'(r)^2-\frac{2k}{r}S_0'(r)-S_0''(r)=2(V(r)-h(r)-g E_0).\eqno(8)
$$
Therefore
$$
h(r)+gE_0=V(r)-\frac{1}{2}(S_0'(r)^2-\frac{2k}{r}S_0'(r)-S_0''(r)).\eqno(9)
$$
For a finite $h(r)$ it should not include terms with positive power
of $r$. Since the highest order of $r$-power in the potential is $6$
and $V(r)$ has only even powers of $r$, we first assume
$$
S_0(r)=(ar^4+cr^2+e)+m\log(\alpha r^2+1).\eqno(10)
$$
Substituting (10) into (8), to cancel $r^6$ term we have $a=g/4$.
Coefficients $e$ and $\alpha$ only change the normalization and we
simply set $e=0$, $\alpha=1$ and finally obtain
$$
S_0(r)=(\frac{g}{4}r^4+cr^2)+m\log(r^2+1).\eqno(11)
$$
To cancel the terms with $r^4$ and $r^2$ we set
$$
c=\frac{1}{4}g(A-2)r_0^2~~{\sf and}~~
m=\frac{1}{4}(g+3)r_0^4-\frac{1}{16}g(A+2)^2r_0^4\eqno(12)
$$
and obtain
$$
h(r)=2m(m+1)\frac{1}{(r^2+1)^2}+(mg(A-2)r_0^2-2mg+2mk-2m^2-m)\frac{1}{r^2+1}\eqno(13)
$$
$$
gE_0=\frac{1}{2}Ag^2r_0^6+2mg
+(2k+1-4m)\frac{1}{4}g(A-2)r_0^2.\eqno(14)
$$
When $g=1$ and $A=2$ we have $m=0$, $h(r)=0$ and the trial
function is just the exact solution of the Schroedinger equation.
To look in more details the behavior of the trial function we
choose $g=1$ and see the change of the trial function with the
parameter $A$. When $A=2$ our trial function is just the exact
solution with the maximum of the wave function at $r=0$. For
$A\neq 2$ we always have $h(r)<0$ and finite, and $h(r)\rightarrow
0$ when $r\rightarrow \infty$. This ensures the convergency of the
iterative procedure. When $A<2$ the potential is more centered at
$r=0$ and the trial function keeps its maximum at $r=0$. When
$A>2$ the potential is more like a double-well and the trial
function has maxima at $r\neq 0$.\\

\noindent{\large \bf Trial Function II}

we can introduce another trial function
$$
\phi(r)=\bigg(\frac{r_0+a}{r+a}\bigg)^k e^{-g{\cal S}_0(r)-{\cal
S}_1(r)}\eqno(15)
$$
satisfying the Schroedinger equation (4) and the boundary condition
$$
\phi(\infty)=0.
$$
The parameter $a$ in (15) is fixed to ensure the boundary condition
$$
\phi'(0)=0,
$$
namely
$$
g{\cal S}_0'(0)+{\cal S}_1'(0)+\frac{k}{a}=0.\eqno(16)
$$
Substituting (15) into (4), we compare terms with the same power of
$g$. From $g^2$-terms we obtain
$$
{\cal S}_0'(r)=\sqrt{2v}=(r^2-r_0^2)\sqrt{r^2+A r_0^2}.\eqno(17)
$$
To ensure the $h(r)$ satisfying the convergence condition, $S_1(r)$
is defined in a special way to prevent terms with positive powers of
$r$ presenting in $h(r)$. For $g^1$ terms we have
$$
-\frac{1}{2}\bigg(2{\cal S}_0'{\cal S}_1'-{\cal
S}_0''-2ka\sqrt{r^2+Ar_0^2}\bigg)\bigg|_{r=r_0}-ka^2=E_0.\eqno(18)
$$
Introducing
$$
E_0=E_0^{(1)}+E_0^{(2)}+E_0^{(3)}\eqno(19)
$$
and defining
$$
E_0^{(3)}=-ka^2\eqno(20)
$$
we write
$$
{\cal S}_0'{\cal S}_1'=(\frac{1}{2}{\cal
S}_0''-E_0^{(1)})+(ka\sqrt{r^2+Ar_0^2}-E_0^{(2)}).
$$
Since $S_0'(r_0)=0$ we obtain
$$
E_0^{(1)}=\frac{1}{2}{\cal S}_0''(r_0)=r_0^2\sqrt{1+A},\eqno(21)
$$
$$
E_0^{(2)}=kar_0\sqrt{1+A}\eqno(22)
$$
and
$$
{\cal S}_1'=(\frac{1}{2}{\cal S}_0''-E_0^{(1)})/{\cal S}_0'
+(ka\sqrt{r^2+Ar_0^2}-E_0^{(2)})/{\cal S}_0'.\eqno(23)
$$
Substituting ${\cal S}_0'(r)$ into (23) we have explicitly
$$
{\cal
S}_1'(r)=\frac{r^2+(1+A)r_0^2}{\sqrt{r^2+Ar_0^2}(r\sqrt{r^2+Ar_0^2}+r_0^2\sqrt{1+A})}
+\frac{r} {2(r^2+Ar_0^2)}
$$
$$
+\frac{ka}{\sqrt{r^2+Ar_0^2}(\sqrt{r^2+Ar_0^2}+r_0\sqrt{1+A})}.\eqno(24)
$$
The expression for $h(r)$ is
$$
h(r)=\frac{1}{2}({\cal S}_1'^2-{\cal
S}_1'')+\frac{1}{2}~\frac{k(k+1)}{(r+a)^2} -\frac{ka}{r(r+a)}{\cal
S}_1'-\frac{k^2}{r(r+a)}
$$
$$
+kag(r_0^2-a^2)\frac{\sqrt{r^2+Ar_0^2}}{r(r+a)}
+ka^2g\frac{Ar_0^2}{r(\sqrt{r^2+Ar_0^2}+r)}.\eqno(25)
$$
Substituting $S_0'(0)$ and $S_1'(0)$ into (16) we obtain an equation
for the parameter $a$
$$
ka^2+(r_0\sqrt{1+A}-gr_0^5A)(\sqrt{A}+\sqrt{1+A})a+kr_0^2(A+\sqrt{A(1+A)})=0.\eqno(26)
$$
For the above equation to have real solutions of $a$ the following
restriction is put on the parameters $g$ and $A$:
$$
(\sqrt{1+A}-gAr_0^4)^2r_0^2(\sqrt{A}+\sqrt{1+A})-4\sqrt{A}k^2r_0^2
\geq 0.\eqno(27)
$$
For example, (27) requires $g>0.922$ when $A=2$ and $A>1.81$ when
$g=1$. When (27) can not be fulfilled the condition $ \phi'(0)=0$
can be satisfied by introducing the trial function as
$$
\phi_{rev}(r)=\phi(r)+\xi \phi_-(r)~~{\sf for}~~r<r_0\eqno(28a)
$$
and
$$
\phi_{rev}(r)=(1+\xi \phi_-(r_0)/\phi(r_0))\phi(r)~~{\sf
for}~~r>r_0\eqno(28b)
$$
where $\phi_-(r)$ is defined as
$$
\phi_-(r)=\bigg(\frac{r_0+a}{r+a}\bigg)^k e^{-g{\cal
S}_0(-r)-{\cal S}_1(r)}.\eqno(29)
$$
The parameter $\xi$ is fixed to satisfy the condition
$\phi_{rev}'(0)=0$, namely
$$
\phi'(0)+\xi \phi'_-(0)=0.\eqno(30)
$$
Correspondingly the Schroedinger equation satisfied by
$\phi_{rev}(r)$ is
$$
(-\frac{1}{r^{2k}}\frac{d}{dr}r^{2k}\frac{d}{dr}+V(r)-h_{rev}(r))\phi_{rev}(r)
=gE_0\phi_{rev}(r),\eqno(31)
$$
where $h_{rev}(r)=h(r)$ for $r>r_0$ and
$$
h_{rev}(r)=h(r)+2g\xi\bigg(E_0+ka\frac{a^2-r_0^2}{r(r+a)}\sqrt{r^2+Ar_0^2}
$$
$$
-ka^2\frac{Ar_0^2}{r(\sqrt{r^2+Ar_0^2}+r)}\bigg)\phi_-(r)/\phi_{rev}(r)\eqno(32)
$$
for $r<r_0$. It is interesting to notice that the conditions (26)
and (30) for $\phi'(0)=0$ also ensure $h(r)$ and $h_{rev}(r)$ to
be finite when $r\rightarrow 0$, which is necessary for the
convergency of the iteration procedure.

By integrating (17) and (25) we obtain ${\cal S}_0(r)$ and ${\cal
S}_1(r)$ as
$$
{\cal S}_0(r)=\frac{1}{8}r\sqrt{r^2+Ar_0^2}(2r^2+Ar_0^2-4r_0^2)
-\frac{1}{8}(A^2r_0^4+4Ar_0^4)\ln (r+\sqrt{r^2+Ar_0^2})\eqno(33)
$$
$$
{\cal S}_1(r)=\ln (r+r_0)+\frac{1}{4}\ln
(r^2+Ar_0^2)+(\frac{1}{2}+\frac{ka}{2r_0}) \ln
\frac{\sqrt{1+A}\sqrt{r^2+Ar_0^2}+r+Ar_0}
{\sqrt{1+A}\sqrt{r^2+Ar_0^2}-r+Ar_0}.\eqno(34)
$$
Substituting them into (15) or (28)-(29) gives the final
expression of the trial functions. From (25) we can also reach
$$
\frac{1}{2}({\cal S}_1'^2-{\cal
S}_1'')=\frac{\gamma}{8(r^2+Ar_0^2)^2(\alpha+\beta)}
+\frac{ka}{2r(r+a)(r^2+Ar_0^2)}\frac{\gamma'}{\alpha'+\beta'}
$$
$$
+\frac{k^2a^2}{2(r^2+Ar_0^2)(\sqrt{r^2+Ar_0^2}+r_0\sqrt{1+A})^2}
$$
$$
+\frac{kar(2\sqrt{r^2+Ar_0^2}+r_0\sqrt{1+A})}{(r^2+Ar_0^2)^{3/2}(\sqrt{r^2+Ar_0^2}+r_0\sqrt{1+A})^2}
\eqno(35)
$$
with
$$
\gamma=225r^8+270(1+2A)r^6r_0^2+3(188A^2+216A-5)r^4r_0^4
$$
$$
+36A(8A^2+10A-1)r^2r_0^6+4A^2(4A+1)^2r_0^8,
$$
$$
\gamma'=9r^4+3(4A+1)r^2r_0^2+4A(1+A)r_0^4,
$$
$$
\alpha=15r^6+(18A-6)r^4r_0^2+(8A^2+12A+7)r^2r_0^4+(8A^2+2A)r_0^6,
$$
$$
\beta=8\sqrt{1+A}r_0^2r\bigg(3r^2+(2A-1)r_0^2\bigg)\sqrt{r^2+Ar_0^2}
$$
and
$$
\alpha'=r(3r^2+(2A-1)r_0^2),
$$
$$
\beta'=2r_0^2\sqrt{1+A}\sqrt{r^2+Ar_0^2}.
$$
Substituting (35) into (25) and (32) gives the final expressions of
$h$ and $h_{rev}$. With above results for $h$, $\phi$ and $h_{rev}$,
$\phi_{rev}$ we are ready to perform the iteration procedure.

\section*{\bf 3. Numerical Result}
\setcounter{section}{3} \setcounter{equation}{0}

Starting from the above defined two sets of trial functions
$\phi(r)$ and the related $h(r)$, we can perform the iterations
based on $f$-iteration of (6a) and (6b) or $\tau$-iteration of
(6c)and (6d). Our numerical results show that although the two
iteration procedures look quite different and the two trial
functions are defined in different ways the finally obtained wave
functions and eigenvalues for the groundstate convergent nicely to
the same final shapes and values. Now we give some more detailed
discussions about our results. Let us take $N=3$ as an example.\\

\noindent {\bf For $g=1$ and $A=2$}

For the trial function I, when $g=1$ and $A=2$, as mentioned before,
the trial function gives just the exact solution of the groundstate
$$
\phi(x)=e^{-r^4/4}
$$
with $E_0=r_0^6$. However, for the trial function II, it is
necessary to fix the parameter $a$ in the trial function (15) first
by solving (26). It gives
$$
a=4.4267~~{\sf or}~~a=1.2976
$$
Performing the iteration based either on $f$-iteration or on
$\tau$-iteration the final convergent result of the wave function
and the eigenvalue of the groundstate is consistent to the exact
solution. The trial function and the final exact wave function for
the groundstate is plotted in Fig.~1. It is interesting to observe
the transition of the shape of the wave function for the trial
function with maxima at a finite $r$ to the final convergent one
with only one maximum at $r=0$ after the iteration procedure, as the
exact groundstate wave function should be. This answered the
question raised by R. Jackiw[1] in $N$-dimensional case: Even the
trial function proposed has its maxima at $r>0$ the iteration
procedure would still reach the exact solution of the groundstate
wave function with its only
maximum at $r=0$.\\

\noindent {\bf Comparison of the Two Iteration Procedures}\\

In Table~1 and Table~2 the eigenvalues of the groundstate obtained
from the $\tau$- and $f$-iterations are listed respectively, based
on the two different trial functions for different parameters $g$
and $A$. Comparing the two iteration procedures, it can be seen
that $\tau$-iteration is convergent faster than $f$-iteration. The
numerical calculation takes also less time to reach the convergent
result for $\tau$-iteration. This can be understood by comparing
the formula (6a) and (6b) for $f$-iteration with (6c) and (6d) for
$\tau$-iteration. First, in the formula for the energy correction,
the denominator changes in each order in $f$-iteration while it
needs only to calculate once for the whole $\tau$-iteration
procedure. Besides, one fold less of integration is needed for
each order of iteration in the $\tau$-iteration procedure since it
is related only to $\tau_n'$. These two advantages speed up the
numerical calculation of $\tau$-iteration very much. Although the
two iteration procedures look quite different with different
convergent speed, it is shown clearly in the two tables that the
two iteration procedures do give the same convergent results.\\

\noindent {\bf Comparison of the Two Trial Functions}\\

For the two trial functions, the trial function I is closer to the
groundstate and needs less orders of iteration to reach the exact
result in most cases. This can be seen clearly for the case of $g=1$
and $A=2$. The trial function I gives already the exact solution
while for the trial function II which has maxima at $r>0$ the exact
groundstate wave function with its only maximum at $r=0$ can be
reached only after the iteration. In fact, for different parameters
$g$ and $A$ the trial functions I always have shapes similar to the
exact solution, while the trial functions II differ from the exact
ones in their shapes for $g\leq 1$ and $A\leq 2$. Although the
iteration process for the two trial functions is quite different,
the two iteration procedures with the two sets of trial functions
always reach the same final
results of eigenvalues and groundstate wave functions.\\

\noindent {\bf Change of the Wave Function Shapes with
Parameters}\\

As examples the obtained groundstate wave functions after the
iteration procedure are plotted in Figs.~2 and 3 for $A=2$ and
$g=0.5,~1$ or $2$, and for $g=1$ and $A=1,~2$ or $3$, respectively.
It is interesting to see the transition of the form of the obtained
groundstate wave function from the shape with maximum at $r=0$ to
the one with maxima at a finite $r$, becoming a degenerate
groundstate, when $g$ increases from $<1$, passing $1$ to $>1$ for
$A=2$, or when $A$ increases from $<2$, passing $2$ to $>2$ for
$g=1$. The results seem to show that the groundstate wave functions
in the region $g\leq 1$ and $A\leq 2$ have the shape with only one
maximum at $r=0$, while in the region outside the wave functions
become degenerate at a finite $r$. Their maxima move to larger $r$
when the parameters $g$ and $A$ increase further.

\newpage

\section*{\bf Appendix}
\setcounter{section}{5} \setcounter{equation}{0}

\vspace{.5cm}

For a particle with unit mass, moving in an N-dimensional potential
$V({\bf r})$, the ground state wave function $\Psi({\bf r})$
satisfies the following Schroedinger equation:
$$
( -\frac{1}{2} {\bf \nabla}^2 + V({\bf r}))\Psi({\bf r}) = E
\Psi({\bf r}).\eqno(A.1)
$$
Introduce a potential correction $h({\bf r})$ and define another
wave function $\Phi({\bf r})$ satisfying the Schroedinger equation
$$
( -\frac{1}{2} {\bf \nabla}^2 + V({\bf r})-h({\bf r}))\Phi({\bf r})
= ( E-\Delta) \Phi({\bf r}),\eqno(A.2)
$$
where $\Delta$ is the corresponding energy correction. Multiplying
(A.2) on the left by $\Psi({\bf r})$ and (A.1) by $\Phi({\bf r})$,
their difference gives
$$
-\frac{1}{2}{\bf \nabla}(\Phi{\bf \nabla}\Psi-\Psi{\bf
\nabla}\Phi)=-(h-\Delta)\Phi\Psi.\eqno(A.3)
$$
Let
$$
\Psi({\bf r}) = \Phi({\bf r})f({\bf r})=\Phi({\bf r})e^{-\tau ({\bf
r})}.\eqno(A.4)
$$
In Ref.[2], an equation is introduced for $(f({\bf r}),~\Delta)$.
Here according to the same procedure similar results are deduced for
$(\tau({\bf r}),~\Delta)$. For the convenience of comparison we list
both results together in the following. Substituting (A.4) into
(A.3) equations for $(f,~\Delta)$ and $(\tau,~\Delta)$ could be
obtained as following:
$$
-\frac{1}{2}{\bf \nabla}(\Phi^2{\bf
\nabla}f)=(-h+\Delta)\Phi^2f\eqno(A.5)
$$
and
$$
\frac{1}{2}{\bf \nabla}(\Phi^2{\bf
\nabla}\tau)=(-h+\Delta+\frac{1}{2}({\bf
\nabla}\tau)^2)\Phi^2.\eqno(A.6)
$$
The integration over all space for the left hand side of above
equations is zero. This leads to the expressions of the energy
correction related to $f$ and $\tau$, respectively:
$$
\Delta = \frac{\int d{\bf r}~\Phi^2 h~f}{\int d{\bf
r}~\Phi^2f}~,\eqno(A.7)
$$
and
$$
\Delta = \frac{\int d{\bf r}~\Phi^2 [h-\frac{1}{2}({\bf
\nabla}\tau)^2]}{\int d{\bf r}~\Phi^2}~.\eqno(A.8)
$$
Introducing
$$
\Delta_1^i,~\Delta_2^i,~\cdots,~\Delta_n^i,~\cdots,~~~i=f~{\rm
or}~~\tau \eqno(A.9)
$$
and
$$
f_1,~f_2,~\cdots,~f_n,~\cdots~~~~{\rm or}~~~~\tau_1,~\tau
_2,~\cdots,~\tau_n,~\cdots\eqno(A.10)
$$
the two iterative series are defined as
$$
\frac{1}{2}{\bf \nabla}(\Phi^2{\bf
\nabla}\tau_n)=(-h+\Delta_n^\tau+\frac{1}{2}({\bf
\nabla}\tau_{n-1})^2)\Phi^2,\eqno(A.11)
$$
$$
\Delta_n^\tau=\frac{\int d{\bf r}~\Phi^2 [h-\frac{1}{2}({\bf
\nabla}\tau_{n-1})^2]}{\int d{\bf r}~\Phi^2}~.\eqno(A.12)
$$
and
$$
-\frac{1}{2}{\bf \nabla}(\Phi^2{\bf
\nabla}f_n)=(-h+\Delta_n^f)\Phi^2f_{n-1},\eqno(A.13)
$$
$$
\Delta_n^f = \frac{\int d{\bf r}~\Phi^2~h~f_{n-1}} {\int d{\bf
r}~\Phi^2~f_{n-1}}~\eqno(A.14)
$$
For later convenience the iteration series for
$\{\tau_n,~\Delta_n^\tau\}$ defined in equations (A.11) and (A.12),
originally introduced in Ref.[4], is named as the $\tau$-iteration,
while the one for $\{f_n,~\Delta_n^f\}$ given in (A.13) and (A.14),
originally introduced in Ref.[2], is named as $f$-iteration in this
paper. By introducing the external electrostatic charge
distributions
$$
\sigma_n^\tau=(-h+\Delta_n^\tau+\frac{1}{2}({\bf
\nabla}\tau_{n-1})^2)\Phi^2\eqno(A.15)
$$
and
$$
\sigma_n^f=(-h+\Delta_n^f)\Phi^2f_{n-1},\eqno(A.16)
$$
correspondingly also defining displacement electric fields
$$
D_n^\tau=\frac{1}{2}\Phi^2{\bf \nabla}\tau_n\eqno(A.17)
$$
and
$$
D_n^f=-\frac{1}{2}\Phi^2{\bf \nabla}f_n,\eqno(A.18)
$$
where one can define $\kappa=\Phi^2$ corresponding to the dielectric
constant in the usual electrostatic problem. Now equations (A.11)
and (A.13) can be expressed by the Maxwell equations of the
electrostatic analog problem for $\tau$- and $f$-iteration[2,4]:
$$
{\bf \nabla}\cdot D_n^i=\sigma_n^i,~~~~~~{\rm
for}~~i=\tau,~f.\eqno(A.19)
$$
If this electrostatic analog problem (A.19) can be solved
numerically, starting from the initial conditions
$$
\Delta_0^\tau=0,~~~ \tau_0=0
$$
or
$$
\Delta_0^f=0,~~~ f_0=1,
$$
the corrections of the groundstate energy and wave function can be
solved by iteration procedures (A.19) and (A.12) or (A.14). However,
the dielectric constant $\kappa=\Phi^2$ in this problem is not a
constant, but changes with ${\bf r}$. This makes it quite
complicated to solve (A.19). Works along this direction are still in
progress.

For radially symmetric potential $V(r)$ and potential correction
$h(r)$ the problem for solving the groundstate can be simplified
and is related only to the radial variable $r$. By separating the
angular variables[2] (A.1) and (A.2) can be reduced to equations
(2) and (4) for the groundstate radial wave functions $\psi(r)$
and $\phi(r)$. Multiplying (4) on the left by $\psi(r)$ and (2) by
$\phi(r)$, their difference gives
$$
-\frac{1}{2r^{2k}}\frac{d}{dr}(\psi r^{2k}\frac{d}{dr}\phi-\phi
r^{2k}\frac{d}{dr}\psi)=(h-\Delta)\phi\psi.\eqno(A.20)
$$
Let
$$
\psi(r) = \phi(r)f(r)=\phi(r)e^{-\tau (r)}.\eqno(A.21)
$$
The equations for $(f(r),~\Delta)$ and $(\tau(r),~\Delta)$ are
deduced as following
$$
\frac{d}{dr}\bigg[r^{2k}\phi^2\frac{df}{dr}\bigg]=2r^{2k}(h-\Delta)\phi^2f\eqno(A.22)
$$
and
$$
-\frac{d}{dr}\bigg[r^{2k}\phi^2\frac{d\tau}{dr}\bigg]=
2r^{2k}\bigg[h-\Delta-\frac{1}{2}\bigg(\frac{d\tau}{dr}\bigg)^2\bigg]\phi^2.\eqno(A.23)
$$
The integration of the left-hand side of Eqs. (A.22) and (A.23) over
$r=0$ to $\infty$ is zero, which gives the expressions of the energy
correction
$$
\Delta=\frac{\int\limits_0^{\infty}r^{2k}\phi^2(r)
h(r)f(r)dr}{\int\limits_0^{\infty}
r^{2k}\phi^2(r)f(r)dr},\eqno(A.24)
$$
and
$$
\Delta = \frac{\int\limits_0^{\infty}r^{2k}\phi^2(r)~
(h(r)-\frac{1}{2}~(\tau'(r))^2) dr} {\int\limits_0^{\infty}
r^{2k}\phi^2(r)~ dr}\eqno(A.25)
$$
with $\tau'=\frac{d\tau}{dr}$. Introducing the two iterative series
$\{f_n(r)\}$, $\{\Delta_n^f\}$ and $\{\tau_n(r)\}$,
$\{\Delta_n^\tau\}$ with $n=0,~1,~\cdots$ it is easy to obtain the
iteration equations (6a)-(6d) from (A.22)-(A.25).

\vspace{1cm}

\noindent {\bf Acknowledgement }

The author would like to thank Professor T. D. Lee for his
continuous guidance and instruction.\\

\vspace{1cm}

\noindent {\bf References}

1. R. Jackiw, Private communication

2. R. Friedberg, T. D. Lee and W. Q. Zhao, Ann. Phys. 321(2006)1981

~~~~R. Friedberg and T. D. Lee, Ann. Phys. 316(2005)44

3. R. Friedberg, T. D. Lee and W. Q. Zhao, arXiv: 0709.1997,

~~~~Ann. Phys. (2007)doi:10.1016/j.aop.2007.09.006

4. Zhao Wei-Qin, Commun. Theor. Phys. 43(2005)1009

\newpage

 Table 1. Eigenvalues of groundstates for $N=3$ based on
 $\tau$-iteration\\

\begin{tabular}{|c|c|c|c|c|c|c|c|}
  \hline
 $g$~~~~$A$ & Trial func. & $E_0$ & $E_1$ & $E_2$ & $E_3$ & $E_4$ & $E_5$\\
  \hline
0.5~~2~& I & 1.1629 & 1.3978 & 1.3763 & 1.3772 & 1.3773 &\\
\hline
0.5~~2~& II & -0.4300 & 1.3963 & 1.3763 & 1.3773 & 1.3773 &\\
\hline
 0.93~~2~ & I & 2.0237 & 2.0352 & 2.0351 & 2.0351 & 2.0351 &\\
  \hline
 0.93~~2~ & II & 2.0921 & 2.0457 & 2.0355 & 2.0351 & 2.0351 &\\
  \hline
 1~~~~2 & I & 2.1517 & &  & & & \\
  \hline
 1~~~~2 & II & -8.6479 & 2.1523 & 2.1517 & 2.1517 & 2.1517 & \\
  \hline
 2~~~~2 & I & 3.6066 & 4.1140 & 4.1093 & 4.1094 & 4.1094 & \\
  \hline
 2~~~~2 & II & 5.5581 & 4.1362 & 4.1123 & 4.1097 & 4.1094 & \\
  \hline
 1~~~~1 & I & 2.5073 & 1.8400 & 1.8392 & 1.8392 & 1.8392 &  \\
  \hline
 1~~~~1 & II & -2.3537 & 1.8920 & 1.8330 & 1.8394 & 1.8393 & 1.8392 \\
  \hline
 1~~1.9 & I & 2.1225 & 2.1215 & 2.1215 & 2.1215 & 2.1215 & \\
  \hline
 1~~1.9 & II & -3.8095 & 2.1232 & 2.1215 & 2.1215 & 2.1215 &\\
  \hline
  1~~~3 & I & 3.5310 & 2.4630 & 2.4426 & 2.4418 & 2.4418 & \\
  \hline
  1~~~3 & II & 3.6773 & 2.4675 & 2.4437 & 2.4419 & 2.4418 &\\
  \hline
\end{tabular}

\newpage

Table 2. Eigenvalues of groundstates for $N=3$ based on
 $f$-iteration\\

\begin{tabular}{|c|c|c|c|c|c|c|c|}
  \hline
 $g$~~~~$A$ & Trial func. & $E_0$ & $E_1$ & $E_2$ & $E_3$ & $E_4$ &$E_5$\\
  \hline
 0.5~~2~& I & 1.1629 & 1.3978 & 1.3705 & 1.3786 & 1.3770 & 1.3773 \\
\hline
 0.5~~2~& II & -0.4300 & 1.3963 & 1.3795 & 1.3775 & 1.3773 & 1.3773 \\
\hline
 0.93~~2~ & I & 2.0237 & 2.0352 & 2.0351 & 2.0351 & 2.0351 &\\
  \hline
 0.93~~2~ & II & 2.0921 & 2.0457 & 2.0337 & 2.0352 & 2.0351 & 2.0351 \\
  \hline
 1~~~~2 & I & 2.1517 & &  &  & & \\
  \hline
 1~~~~2 & II & -8.6479 & 2.1523 & 2.1516 & 2.1517 & 2.1517 & 2.1517 \\
  \hline
 2~~~~2 & I & 3.6066 & 4.1140 & 4.1088 & 4.1094 & 4.1094 & \\
  \hline
 2~~~~2 & II & 5.5581 & 4.1362 & 4.0976 & 4.1108 & 4.1092 & 4.1094\\
  \hline
1~~~1& I & 2.5073 & 1.8400 & 1.8392 & 1.8392 & 1.8392 & \\
\hline
 1~~~1 & II & -2.3537 & 1.8920 & 1.8473 & 1.8402 & 1.8393 & 1.8392 \\
\hline
 1~~1.9 & I & 2.1225 & 2.1215 & 2.1215 & 2.1215 & 2.1215 &  \\
  \hline
 1~~1.9 & II & -3.8095 & 2.1232 & 2.1214 & 2.1215 & 2.1215 & 2.1215\\
  \hline
  1~~~3 & I & 3.5310 & 2.4630 & 2.4464 & 2.4425 & 2.4419 & 2.4418 \\
  \hline
  1~~~3 & II & 3.6773 & 2.4675 & 2.4353 & 2.4425 & 2.4417 & 2.4418 \\
  \hline
\end{tabular}

\newpage

\noindent {\bf Figure Caption}\\

\noindent Fig.~1 Trial Function $\phi(r)$ and Groundstate Wave
 Function $\psi(r)$

 ~~~~~ for $N=3$, $g=1$ and $A=2$.\\

 \noindent Fig.~2 Groundstate Wave
 Function $\psi(r)$ for $N=3$, $g=1$

 ~~~~~ and $A=1$ (thin), $2$ (middle) and $3$
 (thick).\\

\noindent Fig.~3 Groundstate Wave
 Function $\psi(r)$ for $N=3$, $A=2$

 ~~~~~ and $g=0.5$ (thin), $1$ (middle) and $2$
 (thick).\\

\begin{figure}[h]
 \centerline{
\epsfig{file=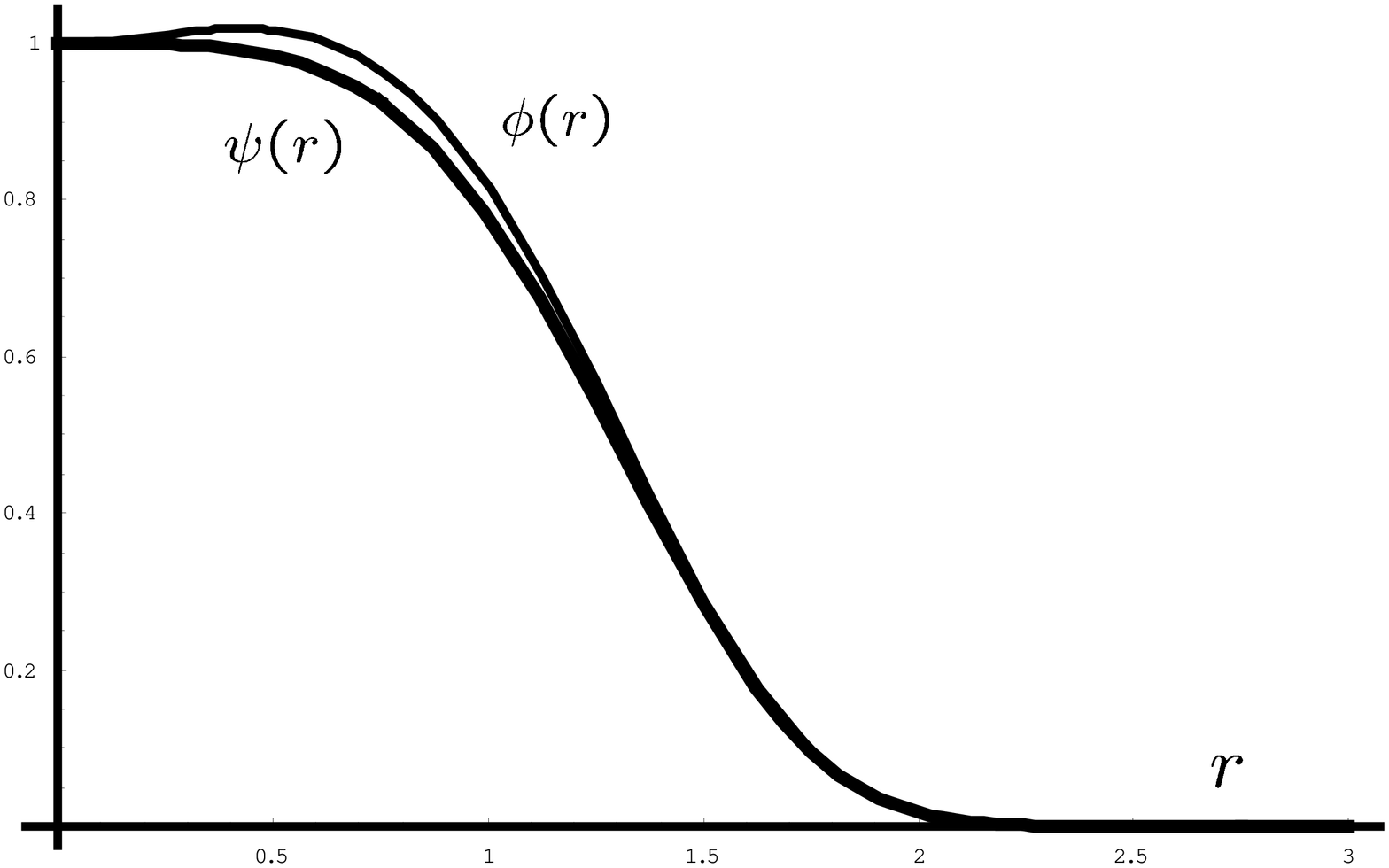, width=8cm, height=14cm, angle=-90}}
\vspace{.5cm}
 \centerline{{\normalsize \sf Fig. 1~  Trial Function $\phi(r)$ and Groundstate Wave
 Function $\psi(r)$ }}
\centerline{{\normalsize \sf for $N=3$, $g=1$ and $A=2$.}}
\end{figure}

\begin{figure}[h]
 \centerline{
\epsfig{file=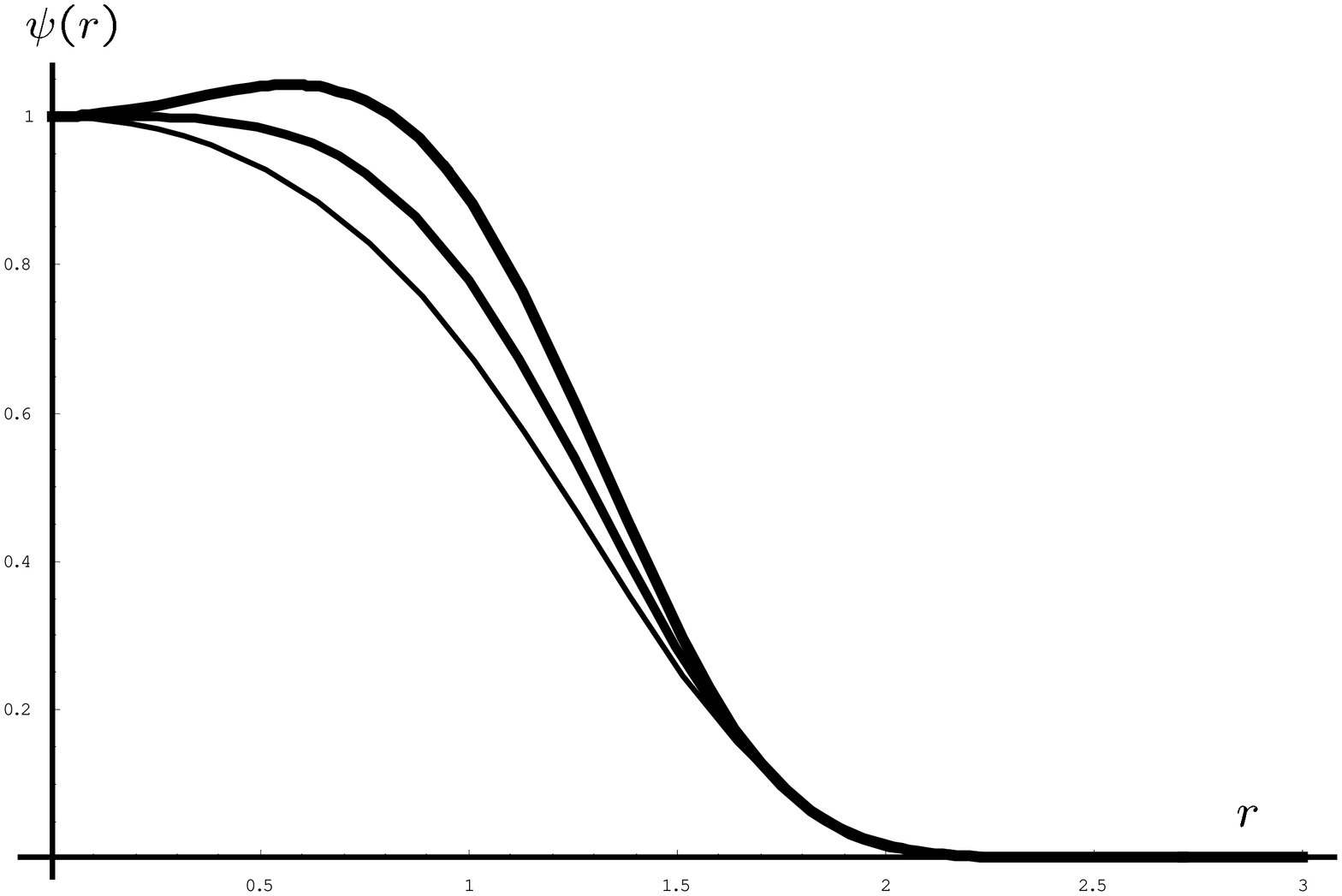, width=8cm, height=14cm, angle=-90}}
\vspace{.5cm}
 \centerline{{\normalsize \sf Fig. 2~  Groundstate Wave
 Function $\psi(r)$ for $N=3$, $g=1$ }}
 \centerline{{\normalsize \sf and $A=1$ (thin), $2$ (middle) and $3$ (thick).}}
\end{figure}

\begin{figure}[h]
 \centerline{
\epsfig{file=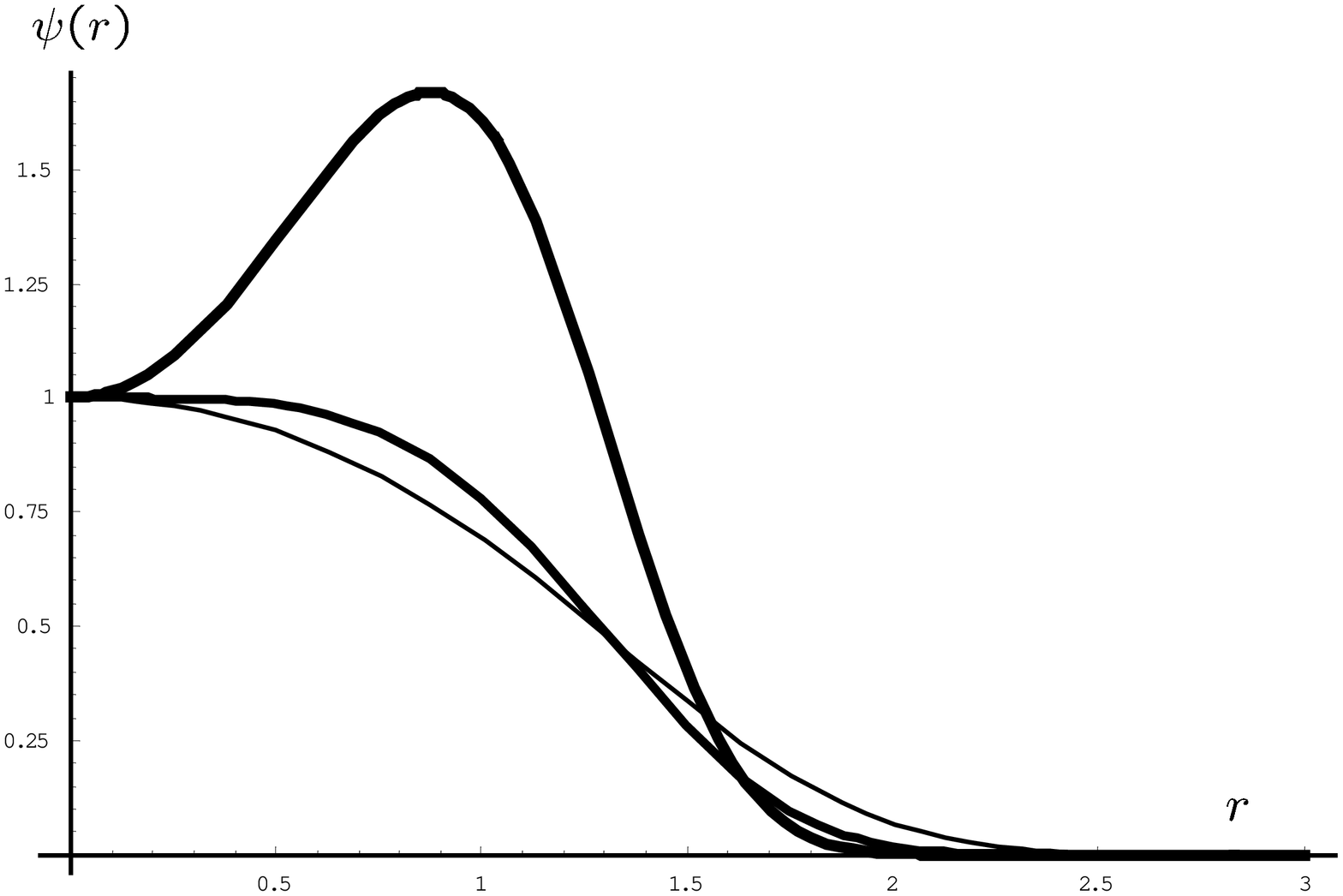, width=8cm, height=14cm, angle=-90}}
\vspace{.5cm}
 \centerline{{\normalsize \sf Fig. 3~  Groundstate Wave
 Function $\psi(r)$ for $N=3$, $A=2$ }}
 \centerline{{\normalsize \sf and $g=0.5$ (thin), $1$ (middle) and $2$ (thick).}}
\end{figure}

 }
\end{document}